\documentstyle[12pt]{article}

\catcode`\@=11
\long\def\@makefntext#1{
\protect\noindent \hbox to 3.2pt {\hskip-.9pt
$^{{\ninerm\@thefnmark}}$\hfil}#1\hfill}		

\def\@makefnmark{\hbox to 0pt{$^{\@thefnmark}$\hss}}  

\def\ps@myheadings{\let\@mkboth\@gobbletwo
\def\@oddhead{\hbox{}
\rightmark\hfil\ninerm\thepage}
\def\@oddfoot{}\def\@evenhead{\ninerm\thepage\hfil
\leftmark\hbox{}}\def\@evenfoot{}
\def\sectionmark##1{}\def\subsectionmark##1{}}

\renewcommand{\thefootnote}{\fnsymbol{footnote}}

\newcounter{sectionc}\newcounter{subsectionc}\newcounter{subsubsectionc}
\renewcommand{\section}[1] {\vspace*{0.6cm}\addtocounter{sectionc}{1}
\setcounter{subsectionc}{0}\setcounter{subsubsectionc}{0}\noindent
	{\normalsize\bf\thesectionc. #1}\par\vspace*{0.4cm}}
\renewcommand{\subsection}[1] {\vspace*{0.6cm}\addtocounter{subsectionc}{1}
	\setcounter{subsubsectionc}{0}\noindent
	{\normalsize\it\thesectionc.\thesubsectionc. #1}\par\vspace*{0.4cm}}
\renewcommand{\subsubsection}[1]
{\vspace*{0.6cm}\addtocounter{subsubsectionc}{1}
	\noindent {\normalsize\rm\thesectionc.\thesubsectionc.\thesubsubsectionc.
	#1}\par\vspace*{0.4cm}}

\newcounter{appendixc}
\newcounter{subappendixc}[appendixc]
\newcounter{subsubappendixc}[subappendixc]

\renewcommand{\appendix}[1] {\vspace*{0.6cm}
        \refstepcounter{appendixc}
        \setcounter{figure}{0}
        \setcounter{table}{0}
        \setcounter{equation}{0}
        \renewcommand{\thefigure}{\Alph{appendixc}.\arabic{figure}}
        \renewcommand{\thetable}{\Alph{appendixc}.\arabic{table}}
        \renewcommand{\theappendixc}{\Alph{appendixc}}
        \renewcommand{\theequation}{\Alph{appendixc}.\arabic{equation}}
        \noindent{\bf Appendix \theappendixc #1}\par\vspace*{0.4cm}}

\def\abstracts#1{{

\centering{\begin{minipage}{12.2truecm}\footnotesize\baselineskip=12pt\noindent
	\centerline{\footnotesize ABSTRACT}\vspace*{0.3cm}
	\parindent=0pt #1
	\end{minipage}}\par}}

\renewenvironment{thebibliography}[1]
	{\begin{list}{\arabic{enumi}.}
	{\usecounter{enumi}\setlength{\parsep}{0pt}
\setlength{\leftmargin 1.25cm}{\rightmargin 0pt}
	 \setlength{\itemsep}{0pt} \settowidth
	{\labelwidth}{#1.}\sloppy}}{\end{list}}

\newcounter{itemlistc}
\newcounter{romanlistc}
\newcounter{alphlistc}
\newcounter{arabiclistc}

\newcommand{\fcaption}[1]{
        \refstepcounter{figure}
        \setbox\@tempboxa = \hbox{\footnotesize Fig.~\thefigure. #1}
        \ifdim \wd\@tempboxa > 6in
           {\begin{center}
        \parbox{6in}{\footnotesize\baselineskip=12pt Fig.~\thefigure. #1}
            \end{center}}
        \else
             {\begin{center}
             {\footnotesize Fig.~\thefigure. #1}
              \end{center}}
        \fi}

\newcommand{\tcaption}[1]{
        \refstepcounter{table}
        \setbox\@tempboxa = \hbox{\footnotesize Table~\thetable. #1}
        \ifdim \wd\@tempboxa > 6in
           {\begin{center}
        \parbox{6in}{\footnotesize\baselineskip=12pt Table~\thetable. #1}
            \end{center}}
        \else
             {\begin{center}
             {\footnotesize Table~\thetable. #1}
              \end{center}}
        \fi}

\def\@citex[#1]#2{\if@filesw\immediate\write\@auxout
	{\string\citation{#2}}\fi
\def\@citea{}\@cite{\@for\@citeb:=#2\do
	{\@citea\def\@citea{,}\@ifundefined
	{b@\@citeb}{{\bf ?}\@warning
	{Citation `\@citeb' on page \thepage \space undefined}}
	{\csname b@\@citeb\endcsname}}}{#1}}

\newif\if@cghi
\def\cite{\@cghitrue\@ifnextchar [{\@tempswatrue
	\@citex}{\@tempswafalse\@citex[]}}
\def\citelow{\@cghifalse\@ifnextchar [{\@tempswatrue
	\@citex}{\@tempswafalse\@citex[]}}
\def\@cite#1#2{{$\null^{#1}$\if@tempswa\typeout
	{IJCGA warning: optional citation argument
	ignored: `#2'} \fi}}

 1
 1
 1

\font\ninerm=cmr9

\input{epsf}

\begin{document}



\newcommand{\eps}{\epsilon}
\newcommand{\vareps}{\varepsilon}

\def\beq{\begin{equation}}
\def\eeq{\end{equation}}
\def\beqa{\begin{eqnarray}}
\def\eeqa{\end{eqnarray}}

\def \psihat{\skew2\widehat{\psi}}
\def \psibarhat{\skew2\widehat{\skew2\bar{\psi}}}
\def \psibar{\skew3\bar{\psi}}
\def \Psibar{\bar{\Psi}}
\def \Psihat{\widehat{\Psi}}

\def \Jhat{\skew5\widehat{J}}
\def \Jms{{\cal{J}}}
\def \Jmshat{{\widehat{\Jms}}}
\def \Jslash{\thinspace{\not{\negthinspace J}}}
\def \Jbar{\skew5\bar{\Jms}}

\def \Poff{{\cal{P}}}
\def \Qoff{{\cal{Q}}}
\def \Loff{{\cal{L}}}
\def \half{ \hbox{$1\over2$} }

\def \permsum#1#2{\sum_{{\cal{P}}(#1\ldots #2)}}
\def \dfour#1{  { {d^4{#1}}\over{(2\pi)^4} }\thinspace  }
\def \dn#1{  { {d^d{#1}}\over{(2\pi)^d} }\thinspace  }
\def \li#1{ {\rm Li_{#1}} }

\def \bra#1{ \langle #1 | }
\def \ket#1{ | #1 \rangle }
\def \braket#1#2{ \langle #1 \thinspace\thinspace #2 \rangle }

\def \Qoffslash{\thinspace{\not{\negthinspace\negthinspace\Qoff}}}
\def \Loffslash{\thinspace{\not{\negthinspace\negthinspace\Loff}}}
\def \kslash{\thinspace{\not{\negthinspace k}}}
\def \qslash{\thinspace{\not{\negthinspace q}}}
\def \mslash{\thinspace{\not{\negthinspace\negthinspace m}}}
\def \Lslash{\thinspace{\not{\negthinspace L}}}

\def \amp{{\cal A}}
\def \minus{{-}}

\def \morespace#1{ \noalign{\vskip#1pt} }

\hbox{  }

\vspace{-1.1in}
\noindent
\hfill Fermilab--Conf--94/421-T

\noindent
\hfill hep-ph/9412350

\vspace{0.10in}

\centerline{\normalsize\bf USE OF RECURSION RELATIONS TO COMPUTE}
\baselineskip=15pt
\centerline{\normalsize\bf ONE-LOOP HELICITY AMPLITUDES\footnote{
To appear in the proceedings of ``Beyond the Standard Model IV'',
held December 13--18, 1994 at Lake Tahoe, CA.}}
\baselineskip=22pt
\centerline{\footnotesize GREGORY MAHLON}
\baselineskip=13pt
\centerline{\footnotesize\it Fermi National Accelerator Laboratory,
M/S 106, P.O. Box 500, Batavia, Illinois 60510}
\baselineskip=12pt
\centerline{\footnotesize E-mail: gdm@fnth03.fnal.gov}

\vspace*{0.35cm}


\abstracts{We illustrate the use of recursion relations in the
computation of certain one-loop helicity amplitudes containing
an arbitrary number of gauge bosons.  After a brief review of the
recursion relations themselves, we discuss the resolution of the
apparent conflict between the spinor helicity method used to solve
the recursion relations and the dimensional regulator used in the
loop integrals.  We then outline the procedure for constructing
loop amplitudes, and present two examples of
results obtained in this manner.}



\vspace*{0.35cm}
\normalsize\baselineskip=14pt
\setcounter{footnote}{0}
\renewcommand{\thefootnote}{\alph{footnote}}

This talk will describe the use of recursion relations
to compute one-loop corrections to multiple gauge boson scattering.
A different approach to the same problem has been discussed
by Dixon at this conference.\cite{DixonsTalk}


We will begin with a brief review of the recursion relation for
the double-off-shell quark current.\cite{BG,QCDloop}  We define the
double-off-shell quark current $\Psihat_{ji}(\Qoff;1,\ldots,n)$
to consist of the sum of all tree diagrams containing exactly one
(massless) quark line with $n$ on-shell gluons attached in
all possible ways.  The quark has momentum
$\Poff$ and color index $i$, the antiquark momentum
$\Qoff$ and color index $j$, and the $\ell$th gluon has
momentum $k_{\ell}$ and color
index $a_{\ell}$.  We take all of the momenta to flow into the
diagram:   $\Poff {+} \Qoff {+} k_1 {+}\cdots {+} k_n = 0$.
Berends and Giele have shown that\cite{BG}
\beq
\hat\Psi_{ji}(\Qoff;1,\ldots,n) = g^n \permsum{1}{n}
[T^{a_1}\cdots T^{a_n}]_{ji}
\Psi(\Qoff;1,\ldots,n)
\eeq
where  $g$ is the gauge coupling, $T^a$ is a representation matrix
for $SU(N)$, $\Psi(\Qoff;1,\ldots,n)$ is the color-ordered current,
and the notation $\Poff(1\ldots n)$ tells us that the sum runs over
all permutations of the gluon labels $\{1,\ldots,n\}$.  The
color-ordered current contains only kinematic factors, and satisfies
the recursion relation
\beq
\Psi(\Qoff;1,\ldots,n) =
- \sum_{j=0}^{n-1}
\Psi(\Qoff;1,\ldots,j)
\Jslash(j{+}1,\ldots,n)
{
{ \Qoffslash {+} \kslash_1 + \cdots + \kslash_n }
\over
{ [\Qoff {+} k_1 {+} \cdots {+} k_n]^2 }
},
\label{RecursionRelation}
\eeq
where $J(j{+}1,\ldots,n)$ is the color-ordered gluon
current, which is derived from the sum of all tree graphs
containing exactly $n{-}j$ on-shell gluons
plus one off-shell gluon.\cite{BG}

The recursion relation (\ref{RecursionRelation}) is easily
solved for the case of $n$ like-helicity gluons using an appropriate
spinor-helicity basis for the gluon polarization vectors.\cite{BG}
Rather than go into the details of the solution, let us note
two of its features.
First, it consists of a sum of terms containing only two propagator
factors each, instead of the maximum of $n$ that might be expected.
Second, for large $\Qoff$, it falls off as $1/\Qoff^3$.  Hence, any
integrals over $\Qoff$ involving just the current and no additional
inverse powers of $\Qoff$ are ultraviolet divergent, bringing
up the question of an appropriate regulator.


The (apparent) difficulty may be summarized as follows:  the solution
of the recursion relation relies heavily upon the use of a spinor
helicity basis for the gluon polarization vectors.  Thus,
the chiral projectors $\half(1\pm\gamma_5)$ play an important role.
Unfortunately, the only viable regulator for QCD is
dimensional regularization, which does not allow for a consistent
definition of $\gamma_5$.  Fortunately, it is possible to separate
the problem into two pieces in such a way that a $d$-dimensional
definition of $\gamma_5$ is not required.

The variant of dimensional regularization which we employ is
the original implementation by 't Hooft and Veltman\cite{regulator}
in which {\it only} the loop momentum is continued to $d$ dimensions;
all external quantities remain in four dimensions.  Denoting
the $d$-dimensional loop momentum by $\Loff$, we may write
$ \Loff \equiv L + m $,
where $L$ contains only the usual four space-time components of the
momentum, and $m$ contains only the ``extra'' components generated
by the continuation to $d$ dimensions.  Setting $m^2 \equiv -\mu^2$,
this decomposition implies that
$L\cdot m =0$, $\Loff^2 = L^2 - \mu^2,$ and
$\Loff \cdot q = L\cdot q$, where $q$ is any external vector.
Furthermore, $\mslash$ anticommutes with $\qslash$ and $\Lslash$
(but not $\Loffslash$).  At one loop, it is clear that the momentum
shifts required to do the integration involve only the first four
components of $\Loff$.  Hence, any term containing an odd number of
factors of $m$ integrates to zero.

Based on these properties, we can  take any
expression we would have written for a particular amplitude using
the traditional Feynman rules and rewrite it as a sum of terms
whose numerator dependence on the loop momentum has the form
$L^{\nu_1} \cdots L^{\nu_i} \mu^{2j}$.  Since only four-dimensional
vectors are contracted into the Dirac matrices in this
form,  we may translate these expressions into
spinor helicity notation and solve the recursion relation
as usual. The only difference is that we must keep track
of any additional $\mu^2$-containing
terms which may be generated along the way.\footnote{
Such terms are generated because we do not expand the propagators
in powers of $\mu^2$.  Hence, the numerators are effectively
four-dimensional, while the denominators are $d$-dimensional.
Accordingly, propagators are ``cancelled'' using  the relation
$ L^2/\Loff^2 = 1 + \mu^2/\Loff^2$.}
This procedure is only
practical if many of these ``extra'' terms vanish, and there is
a simple prescription for identifying those terms which do not
vanish.

Consider the loop integral which contains $m$ powers of $\mu^2$
and $n$ denominators:
\beq
{\cal J} \equiv
\int \negthinspace\dn{\Loff}
{
{\mu^{2m}}
\over
{ \Loff^2 [\Loff{+}q_1]^2 [\Loff{+}q_2]^2 \cdots [\Loff{+}q_{n-1}]^2 }
}.
\label{sample}
\eeq
{}From the discussion below, it will be obvious how to handle
the case including powers of $L$.
We may introduce Feynman parameters and carry out the momentum
integration by writing $d^d\Loff = d^4L d^{d{-}4}m$, with the
result
\beq
{\cal J} =
-{ {i(-1)^n}
   \over
   {16\pi^2} }
{ { {\eps}(4\pi)^{\eps} }
   \over
   { \Gamma(1{-}\eps) } }
\Gamma(m{-}\eps)
\Gamma(n{-}m{-}2{+}\eps)
\int_0^{\infty} \negthinspace\negthinspace\negthinspace
d^{n}z \thinspace\delta(1-\Sigma z)
{ \bigl[- f(q_i,z_i)\bigr]^{2+m-n-\eps} }.
\label{KeyIntegral}
\eeq
Here $f(q_i,z_i)$ is a function of the external momenta and
Feynman parameters, and $\eps = (d{-}4)/2$.
Note that (\ref{KeyIntegral}) contains an overall factor of $\eps$.
Hence, the expression will vanish in the limit
$\eps\rightarrow 0$ unless one of the other factors generates
a pole in $\eps$.

The first source of such a pole is the factor $\Gamma(m{-}\eps)$,
which is singular if \hbox{$m=0$.}  This term, which has no powers of
$\mu^2$, is in some sense the ``leading'' term:  it corresponds
to the four-dimensional part of the numerator.

If $m\geq n{-}2$, then the factor $\Gamma(n{-}m{-}2{+}\eps)$
produces a pole.  Power counting applied to such integrals
reveals that they are ultraviolet divergent in four
dimensions ($\mu^2$ counts as
two powers of loop momentum, just like $L^2$).
Such terms occur only in the last stages of solution of the
recursion relation, or in certain graphs with only a few legs
attached to the loop.  Furthermore, if $m=n{-}2$, the parameter
integral is trivial to compute.  This case comes up quite often,
and the result is simple.\cite{SimpleIntegral}

The final potential source of poles in $\eps$ is the parameter
integral. These are only infrared in nature.
The exponent of $-f(q_i,z_i)$  is the
one corresponding to a theory in $4{+}2m$ dimensions, where
logarithmic infrared divergences disappear ($m\geq1$).
Thus,
in the vast majority of cases,
the parameter integral is finite.

We may summarize the above by the following two statements.
First, integrals containing no numerator powers of $\mu^2$
must always be computed.  Second, integrals containing one or
more powers of $\mu^2$ contribute
only if they would be ultraviolet divergent in four dimensions.


Armed with this knowledge, we are ready to proceed to a description
of amplitude building based upon the recursion relation solutions.
The first step is to ``glue'' one or more currents together with
one or more of the vertices of the theory.  For example, all of the
contributions to the $n$-gluon amplitude that proceed through a
quark loop are represented by Fig. 1, which consists of a
double-off-shell quark current and a gluon current tied together
at a single $q \bar q g$  vertex.  Notice that in order to obtain
all of the diagrams of this type, we must perform a sum over all
of the ways to divide the $n$ gluons between the two currents.  Thus,
the next step in the procedure is to perform as much of this
algebra as possible.  It is beneficial to watch for and take advantage
of opportunities to reduce the number of propagators in each of
the terms of the expression.

\includegraphics{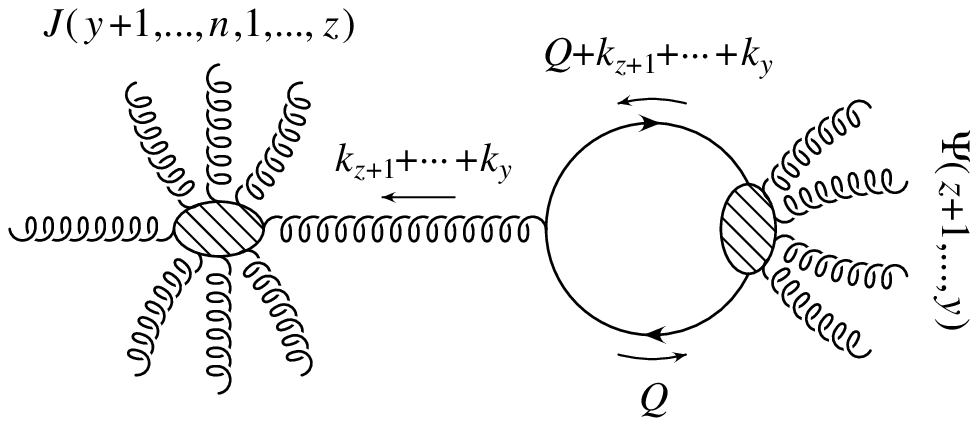}
\begin{figure}
\vskip1.75in
\fcaption{Contributions to the
$n$-gluon scattering amplitude involving a quark loop.}
\vskip-0.15in
\end{figure}

Once the contributions to the integrand are as simple as possible,
we perform the momentum integration.  The shift in loop momentum
required for each term is determined and applied to the numerators.
The parameter integrals are then performed
using the differentiation method of
Bern, Dixon and Kosower,\cite{TensorIntegrals}  which allows
integrals containing extra numerator factors of the Feynman parameters
to be written as derivatives of the corresponding scalar integral.
The differentiation process is simple enough to
let us prepare a table of such integrals.  Given this table,
it is straightforward to cancel any spurious divergences introduced
in the reduction process.  At this stage, all that
remains is to ``clean up'' the result.


We now present two examples of results obtained by the above
procedure.  The first is for the process $\gamma\gamma\rightarrow
\gamma\cdots\gamma$, for the helicity configuration $(\minus\minus
{+}{+}\cdots{+})$.\footnote{The helicity labels are always those
for inward-directed momenta.}  This calculation
involves at worst box integrals---no higher
point functions appear, even for an arbitrary number of photons.
The result for even $n\geq 6$ reads
\beq
\amp(1^{-},2^{-},3^{+},\ldots,n^{+})  =
- { {i(-e\sqrt2)^n}\over{8\pi^2} }
\negthinspace\negthinspace\negthinspace
\permsum{3}{n} \sum_{j=4}^n
{
{ [  \braket{1}{3}
    {\braket{j}{3}}^{*}
     \braket{j}{2}  ]^2 }
\over
{\braket{3}{4} \braket{4}{5} \cdots \braket{n}{3} }
}
{
{ \Lambda(3,\ldots,j) }
\over
{ (2k_3\cdot k_j)^2 }
},
\label{TwoMinus}
\eeq
where the spinor inner products $\braket{i}{j}$
satisfy $\braket{i}{j} { \braket{i}{j} }^{*} = 2k_i \cdot k_j$.
The function $\Lambda(3,\ldots,j)$ is a particular combination
of dilogarithms given by
\beqa
\Lambda(3,\ldots,j) &=&
\li{2}\biggl[ 1{-}
{ { q^2(3,j) q^2(4,j{-}1) } \over {q^2(4,j) q^2(3,j{-}1)} }
\biggr]
- \li{2}\biggl[ 1{-}
{ { q^2(3,j)} \over {q^2(4,j)} }
\biggr]
- \li{2}\biggl[ 1{-}
{ { q^2(4,j{-}1)} \over {q^2(3,j{-}1)} }
\biggr]
\cr\morespace{2}
\negthinspace &-& \negthinspace \li{2}\Biggl[ 1{-}
{ { q^2(3,j)} \over {q^2(3,j{-}1)} }
\biggr]
- \li{2}\biggl[ 1{-}
{ { q^2(4,j{-}1)} \over {q^2(4,j)} }
\biggr]
- {{1}\over{2}} \ln^{2}\biggl[
{ { q^2(3,j{-}1)} \over {q^2(4,j)} }
\biggr],
\eeqa
with $q(i,j) \equiv k_2{+}k_i{+}\cdots{+}k_j$.  Although it is not
immediately obvious, Eq. \ref{TwoMinus} is indeed symmetric under
the interchange $k_1 \leftrightarrow k_2$, as dictated by Bose
statistics.

The second example is the expression for the quark-loop
contributions to the scattering of $n$ like-helicity
gluons.\cite{QCDloop}  We find
\beq
{\cal{A}}(1^{+},\ldots,n^{+}) =
-{
{i(-g\sqrt2)^n}
\over
{48\pi^2}
}
\negthinspace\negthinspace\negthinspace
\permsum{1}{n{-}1}
\negthinspace\negthinspace\negthinspace
tr(T^{a_1}\cdots T^{a_n})
\sum_{j=2}^{n-2}
\sum_{\ell=j{+}1}^{n-1}
{
{ {\rm Tr}\{ \bar k_j \kappa(1,j) \bar\kappa(1,\ell) k_{\ell} \} }
\over
{ \braket{1}{2}\braket{2}{3} \cdots \braket{n}{1} }
},
\label{nGluons}
\eeq
where $\kappa(1,j) \equiv k_1 {+} \cdots {+} k_j$.

In this talk, we have illustrated the use of recursion relations
to simplify the computation of certain
loop amplitudes containing an
arbitrary number of external gauge bosons.  Although there
is still much work to be done before next-to-leading order
cross sections may be extracted from these expressions, a great
deal of progress has been made in this area, and real predictions
are in sight on the horizon.


\vspace*{0.5cm} \noindent {\normalsize\bf References}

\end{document}